# Demixing and tetratic ordering in some binary mixtures of hard superellipses

Sakine Mizani[1], Péter Gurin[2], Roohollah Aliabadi[3], Hamdollah Salehi[1] and Szabolcs Varga[2]

[1]Department of Physics, Faculty of Science, Shahid Chamran University of Ahvaz, Ahvaz, Iran

[2] Institute of Physics and Mechatronics, University of Pannonia, P.O. Box 158, Veszprém, H-8201, Hungary

[3]Department of Physics, Faculty of Science, Fasa University, 74617-81189 Fasa, Iran

## Abstract

We examine the fluid phase behaviour of the binary mixture of hard superellipses using the scaled particle theory. The superellipse is a general two-dimensional convex object, which can be tuned between circular and rectangular shapes continuously at a given aspect ratio. We find that the shape of the particle affects strongly the stability of isotropic, nematic and tetratic phases even if the aspect ratios of both species are fixed. While the isotropic–isotropic demixing transition can be ruled out using the scaled particle theory, the first order isotropic-nematic and the nematic–nematic demixing transition can be stabilized with strong fractionation between the components. It is observed that the demixing tendency is strongest in small rectangle–large ellipse mixtures. Interestingly, it is possible to stabilize the tetratic order at lower densities in the mixture of hard squares and rectangles where the long rectangles form nematic phase, while the squares stay in tetratic order.

# Introduction

With the sudden development of the new colloidal and granular materials, the tailoring technique of the nano- and microparticles have created new shapes of particles (e.g. superballs, lenses, stars) which exhibit very interesting percolation, glass formation, jamming behaviour and phase transitions [1-3]. To understand the observed structural and phase behaviours, several geometrical properties (curvature, volume, contact distance, excluded volume) have to be determined since many macroscopic properties depends on the details of the shape and the size of the particles [4].

One of the simplest phase transition of rod-like particles, where the shape really matters is the nematic–smectic A phase transition of rod-like particles. In the system of hard particles, which consists of cylinders or spherocylinders, it is observed that the smectic A phase is stable if the particles are at least moderately elongated [5-6], while the smectic A phase is totally absent in the system of hard ellipsoids [7]. To understand this strange behaviour it was useful to introduce new geometrical models for the particles, which interpolate between the cylindrical and ellipsoidal shapes. In this regard the superellipsoidal [8-9] and the sphero-ellipsoidal [10] shapes have served new information about the importance of the roundness of the central part and the end parts of the particles, respectively. The other example is the suspension of colloidal superballs, where the shape is between spherical and cubic, exhibits richer phase behaviour than those of spheres or cubes. In the suspension of superballs, new solid phases are observed including a solid-solid transition between a plastic and a rhombohedral crystal [11]. The monolayer of colloidal superballs behaves very similarly because a phase transition between a hexagonal rotator crystal phase and a rhombic crystal phase emerges [12]. To explain this phase transition the simplest model is not the two-dimensional (2D) system of hard squares because the hard squares form only tetratic and square crystal phases [13-15]. Avendaño and Escobedo have performed Monte Carlo simulations with a rounded hard-square model, which interpolates between squares and disks, to examine the effect of roundness on the stability of crystal phases. They showed that the roundness of the vertices is crucial in the formation of rotor crystal phase [16]. With changing the shape of the 2D particles the simulation studies and the vibration experiments revealed the existence of new phases such as the triatic phase of triangles, the tetratic phase of squares and the hexatic phase of hexagons, which are intermediate between fluid and solid phases [17-22]. Interestingly the system of hard pentagons does not form intermediate phases between the fluid and solid phases [23]. From these results it is evident that future studies require new geometrical models which can be tuned continuously between different shapes.

The hard superellipse model is a good candidate to interpolate between two limiting shapes. If the lengths of both sides of the superellipse are equal, we get the 2D superball model, which interpolates between a disk and a square. The ellipse and the rectangle are also the limiting shapes of the superellipse, if the side lengths are unequal. The first milestone in the studies of hard superellipses was the exact determination of the maximal packing arrangement and the corresponding maximal packing fraction of hard superdisks [24]. Later, the jamming properties [25] and the kinetics of randomly packed superdisks [26] were also examined. These studies was extended to binary mixtures of hard superdisks to locate the jammed state [27]. Regarding the hard superellipses we are only aware of the contact point calculations [28] and the percolation threshold study of overlapping superellipses [29]. To study the phase behaviour of hard superellipses, the elaboration of overlap check between superellipses is the essential part in the simulations, while the excluded distances and areas are the key quantities in the mean-field theories such as the Onsager-theory and the scaled particle theory. The determination of these quantities are not trivial even for hard ellipses [30,31]. The issues of the excluded area calculation for convex and concave 2D objects are considered in two recent publications [32,33]. In our present study we show that the exclude area between two different superellipses having different sizes and shapes can be determined analytically.

It is a well-known fact that the phase behaviour of mixtures is richer than those of one-component systems if the size and the shape of the components differ substantially. The competing different structures give rise to strong segregation and even demixing between identical phases [34]. In two dimensions the simplest binary mixture, which exhibit demixing transition is the nonadditive mixture of hard disks [35-41]. The positive nonadditivity is responsible for the fluid-fluid demixing transition, because it reduces the available room for both components upon mixing. It is showed that both symmetric (diameters of the components are the same) and non-symmetric hard disk mixtures can demix if the nonadditivity exceeds a minimum value [38,42,43]. The additive 3D hard body mixtures can exhibit isotropic-isotropic and nematic-nematic transitions if the constituting particles are rod-like [44-46]. Opposite to 3D rod-rod mixtures, only demixing transitions in nematic phase is found in binary mixtures of rod–like particles such as the rectangle-rectangle and ellipse–ellipse mixtures [47-50]. If the shape between the particles is not very different like in the mixture of hard disks and hard ellipses, the demixing transition and the segregation are not present, but both rotational and translational glass transitions may occur [51].

In this work we examine the effects of varying shapes and sizes in the phase behavior of two-dimensional hard convex objects using the scaled particle theory. We choose the superellipse shape, which provides a bridge between circular and rectangular shapes. This shape allows us to consider new mixtures such as the ellipse–rectangle, where the shape of ellipse can be tuned into the direction of rectangle and vice versa for rectangle. We concentrate only on the stability of isotropic, nematic and tetratic ordering and search for possible phase transitions such as the isotropic–nematic, isotropic–tetratic and nematic–nematic ones. As the inputs of the theory are the areas of the particles and the excluded area between two particles, we calculate the area of the superellipse and derive an algorithm for the excluded area between two superellipses. To understand the shape dependence of the observed segregations (isotropic-nematic fractionation, nematic–nematic demixing) we make a link between the superellipse mixture and the nonadditive mixture of hard disks, where the nonadditivity parameter is the driving force of the fluid-fluid demixing. We finish our study with presenting a possible stabilization method of the low density tetratic order with mixing hard squares with long rod-like particles.

## Model

In this work we examine the phase behaviour of some binary mixtures of hard 2D objects, where both the shape and size of the particles can be tuned easily. This 2D object is the hard superellipse with a semi-major axis length ($a$) and a semi-minor axis length ($b$). The equation of superellipse depends also on the deformation parameter ($n$) as follows

$$\left(\frac{|x|}{a}\right)^n + \left(\frac{|y|}{b}\right)^n = 1 \tag{1}$$

where $n \geq 2$ guaranties that the superellipse has a smooth curve and its tangent is well defined in its every point. For this reason, we deal only with this case in this paper. With increasing $n$, the shape of the particle can be tuned from the circular shape to the rectangular one. This opens the window to study a wider class of binary mixtures as the lengths and the shapes of both components can be varied independently. Using the diameter of component one ($2b_1$) as a unit of the length, we have the following five independent molecular parameters: I) the aspect ratio of component 1 ($k_1 = a_1/b_1$), II) the aspect ratio of component 2 ($k_2 = a_2/b_2$), III) the diameter ratio ($d = b_2/b_1$) and IV)-V) the exponents ($n_1$ and $n_2$). Binary mixtures, which cannot be studied with the simple ellipse shape, are the followings: the square–square, square–ellipse, ellipse–rectangle and rectangle–rectangle mixtures. Note that this list is even longer if we replace one of the components or both components with superellipses having intermediate values of $n_1$ and $n_2$ which are between 2 (ellipse-limit) and infinity (rectangle-limit). It is also possible to study the binary mixture of hard superdisks if $a_1 = b_1$ and $a_2 = b_2$. Our experience is that we can get back the perfect rectangle shape and the area of the rectangle very accurately if $n = 100$. For this reason $n_1$ and $n_2$ is always between 2 and 100 in this study. The components of one possible binary mixture is shown in Fig. 1, where the first component is ellipse, while the second component is superellipse.

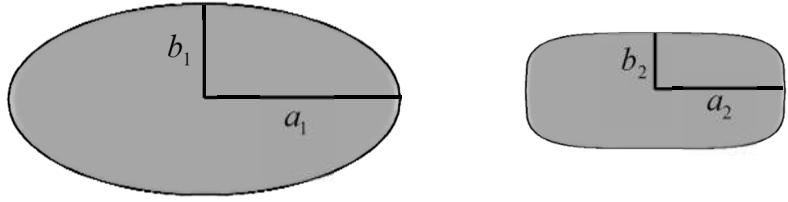

**Figure 1**: The particles of a possible superellipse mixture: a standard ellipse with $n_1 = 2$ (left) and a superellipse with $n_2 = 4$ (right) are shown.

## Theory

We focus on the isotropic, tetratic and nematic phases of the binary mixture using the scaled particle theory (SPT). In the SPT approach [47], the free energy density of the binary mixture can be written as

$$\frac{\beta F}{A} = \sum_{i=1}^{2} \rho_i \left( \ln \rho_i - 1 + \sigma[f_i] \right) - \rho \left( \ln(1-\eta) + \frac{\eta}{1-\eta} \right) + \frac{1}{2(1-\eta)} \sum_{i,j=1}^{2} \rho_i \rho_j \langle\langle A_{exc}^{ij} \rangle\rangle, \qquad (2)$$

where $\beta = 1/k_B T$ is the inverse temperature, $A$ is the area of the surface, $\rho_i = N_i/A$ is the density of the component $i$ ($i = 1, 2$), $N_i$ is the number of particles of component $i$, $\eta = \rho_1 \tilde{a}_1 + \rho_2 \tilde{a}_2$ is the packing fraction, $\tilde{a}_i$ is the area of the particle of component $i$ and $A_{exc}^{ij}$ is the excluded area between a particle of component *i* and a particle of component *j*. In Eq. (2) we use the following short notations for the sake of brevity:

$$\sigma[f_i] = \int_0^{2\pi} f_i(\varphi) \ln f_i(\varphi) d\varphi \tag{3}$$

and

$$\left\langle\left\langle A_{exc}^{ij}\right\rangle\right\rangle = \int_0^{2\pi}\int_0^{2\pi} f_i(\varphi_1) f_j(\varphi_2) A_{exc}^{ij}(\varphi_1 - \varphi_2) d\varphi_1 d\varphi_2, \tag{4}$$

where $f_i$ is the orientational distribution function of the component *i* and $\varphi_i$ is the orientation of a particle of component *i*. In order to get the equilibrium orientational distribution functions, we need to minimize the free energy with respect to $f_i$. This procedure results in a coupled integral equations for $f_i$, which can be written down as

$$f_i(\varphi) = \frac{\exp\left[-\frac{1}{1-\eta}\sum_{j=1,2}\rho_j \int_0^{2\pi} A_{exc}^{ij}(\varphi - \varphi_1) f_j(\varphi_1) d\varphi_1\right]}{\int_0^{2\pi} d\varphi_2 \exp\left[-\frac{1}{1-\eta}\sum_{j=1,2}\rho_j \int_0^{2\pi} A_{exc}^{ij}(\varphi_2 - \varphi_1) f_j(\varphi_1) d\varphi_1\right]} \quad (i=1,2) \tag{5}$$

To solve Eqs. (5), we use nematic and tetratic trial functions for $f_1$ and $f_2$ as initial guesses, we do numerical integrations and use iteration method. After having obtained the equilibrium distribution functions $f_1$ and $f_2$, we measure the extent of the orientational ordering of both components with the two-dimensional order parameters, which are defined as

$$S_2(i) = \int_0^{2\pi} f_i(\varphi)\cos(2\varphi) d\varphi, \tag{6a}$$

$$S_4(i) = \int_0^{2\pi} f_i(\varphi)\cos(4\varphi) d\varphi, \tag{6b}$$

where $S_2(i)$ and $S_4(i)$ measure the nematic and tetratic ordering of the component *i*, respectively. Note that $S_j(i) = \int_0^{2\pi} f_i(\varphi)\cos(j\varphi) d\varphi$ is the general order parameter of component *i*, where $j = 2, 4, 6,...$ etc.

In order to get some information about the orientational ordering transition, we perform bifurcation analysis, which gives the stability limit of the isotropic phase with respect to ordered phases such as the nematic and tetratic phases. In the case of the second order phase transition this analysis provides either the isotropic–nematic or the isotropic–tetratic transition density. To perform this analysis we use the Fourier expansion of the excluded area, which can be written as

$$A_{exc}^{ij}(\varphi_{12}) = \sum_{k=0}^{\infty} A^{ij,k} \cos(k\varphi_{12}) \tag{7}$$

where $A^{ij,k}$ is the Fourier component of the expansion and $\varphi_{12} = \varphi_1 - \varphi_2$. Using Eq. (7) the integrals in Eq. (5) can be written as

$$\int_0^{2\pi} A_{exc}^{ij}(\varphi - \varphi_1) f_j(\varphi_1) d\varphi = \sum_{k=0}^{\infty} A^{ij,k} S_k(j) \cos(k\varphi) \tag{8}$$

In the case of nematic phase $S_2(j)$ start to be nonzero first, while $S_4(j)$ is the first non-vanishing term for the tetratic ordering. Therefore $S_2(j)$ and $S_4(j)$ terms are taken into account at the IN and IT bifurcations, respectively, while the other order parameter terms can be considered to be zero. Denoting the first non-vanishing order parameter as $S_l(j)$ we get that

$$f_i(\varphi) = \frac{\exp\left[-\frac{1}{1-\eta} \sum_{j=1,2} \rho_j A^{ij,l} S_l(j) \cos(l\varphi)\right]}{2\pi} \tag{9}$$

As $S_l(j)$ goes to zero at the bifurcation, we can perform a Taylor-expansion in Eq. (9) as $\exp(-\varsigma) \approx 1 - \varsigma$. In the final step we multiply both sides of Eq. (9) with $\cos(l\varphi)$ and integrate out the $\varphi$ dependence to arrive into two coupled equations for the order parameters. We obtain the following equations

$$\left(1 + \frac{c\rho_1 A^{11,l}}{2}\right) S_l(1) + \left(\frac{c\rho_2 A_{exc}^{12,l}}{2}\right) S_l(2) = 0,$$
$$\left(\frac{c\rho_1 A_{exc}^{21,l}}{2}\right) S_l(1) + \left(1 + \frac{c\rho_2 A_{exc}^{22,l}}{2}\right) S_l(2) = 0, \tag{10}$$

where $c = 1/1-\eta$. These equations can be written down as a product of a matrix and a vector, where the matrix elements are the prefactors of the order parameters, while the order parameters constitute the vector. To have a non-trivial solution of Eq. (10) the determinant of the matrix must be zero. This produces the following bifurcation equation

$$\left(1+\frac{c\rho_1 A^{11,l}}{2}\right)\left(1+\frac{c\rho_2 A^{22,l}}{2}\right)-\frac{c^2\rho_1\rho_2 A^{12,l} A^{21,l}}{4} = 0 \tag{11}$$

We solve this equation numerically for a given molecular parameters ($k_1, k_2, d, n_1,$ and $n_2$) and mole fractions ($x_i = N_i / N$) to get the bifurcation density ($\rho$) and packing fraction ($\eta$) as $\rho_i = x_i \rho$.

In the case of 1$^{st}$ order phase transitions we need to know the chemical potentials of each components and the pressure of the system. These quantities come from the following equations:

$$\mu_i = \frac{\partial\left(\frac{F}{A}\right)}{\partial \rho_i}, \tag{12}$$

$$P = -\frac{F}{A} + \sum_{i=1}^{2} \mu_i \rho_i. \tag{13}$$

Applying above equations we can determine the equation of the state of the system and study the stability of the isotropic, nematic and tetratic phases of the binary mixture. The coexisting densities of a phase transition can be calculated using $\mu_1^\alpha = \mu_1^\beta$, $\mu_2^\alpha = \mu_2^\beta$ and $P^\alpha = P^\beta$, where the phases $\alpha$ and $\beta$ are in equilibrium.

In order to solve Eqs. (5) and (11) at a given density and composition, we need to know the molecular areas and the excluded areas of hard superellipses. It can be proved that

$\tilde{a}_i = \frac{a_i b_i}{n_i} 2^{2\left(1-\frac{1}{n_i}\right)} \sqrt{\pi} \frac{\Gamma\left(\frac{1}{n_i}\right)}{\Gamma\left(\frac{1}{n_i}+\frac{1}{2}\right)}$ is the area of a superellipse of the component $i$, where $\Gamma$ is the

Gamma function. In the ellipse-limit ($n_i = 2$) we can see that $\tilde{a}_i = \pi a_i b_i$, while $\tilde{a}_i = 4 a_i b_i$ is in the rectangle-limit ($n_i \to \infty$). The excluded area between a particle $i$ with orientation $\varphi_i$ and a particle *j* with orientation $\varphi_j$, when the angle between the main axes of these two particles is $\varphi_{ij} = \varphi_i - \varphi_j$, can be written as follows

$$A_{exc}^{ij}(\varphi_{ij}) = \tilde{a}_i + \tilde{a}_j + 2\int_0^\pi \left( \left[ \frac{dr_j}{d\theta_j} \cos(\theta_j + \varphi_{ij}) - r_j \sin(\theta_j + \varphi_{ij}) \right] \sin\theta_i - \left[ \frac{dr_j}{d\theta_j} \sin(\theta_j + \varphi_{ij}) + r_j \cos(\theta_j + \varphi_{ij}) \right] \cos\theta_i \right) r_i d\theta_j \quad (14)$$

where $r_i = \left[ \left( \frac{|\cos\theta_i|}{a_i} \right)^{n_i} + \left( \frac{|\sin\theta_i|}{b_i} \right)^{n_i} \right]^{-\frac{1}{n_i}}$. Note that $\theta_i$ has a complicate dependence from $\theta_j$. Therefore we present the details of the excluded area calculation in the Appendix. As the diameter of component one $(2b_1)$ taken to the unit of the length, $P^* = 4b_1^2 \beta P$ is the dimensionless pressure and $A_{exc}^* = A_{exc}/(2b_1)^2$ is the dimensionless excluded area.

## Results

We start to show how the two-dimensional hard superellipse changes its shape from the ellipse to rectangle with increasing the deformation parameter (*n*). We can see in Fig. 2(a) that the increasing *n* makes the particle more and more rectangular. The case of $n = 2$ corresponds to an ellipse, while $n = 100$ almost to a rectangle. Practically, we find that the area of the rectangle is identical with that of superellipse if $n = 100$. The excluded area between two hard superellipses for all possible combinations of *n* (2, 4, 6 and 100) is shown as a function of $\varphi_{12}$ angle in Fig. 2(b). The effects of increasing $n_1$ and $n_2$ shift the excluded area to higher values and gives rise to a local minimum at $\varphi_{12} = \pi/2$. It is interesting that $n_1 = n_2 = 4$ case is an intermediate shape between ellipse and rectangle as the local minimum just appears at $\varphi_{12} = \pi/2$. We can also see that the ellipse shape is the best for nematic ordering as this shape provides the highest excluded area gain with parallel alignment.

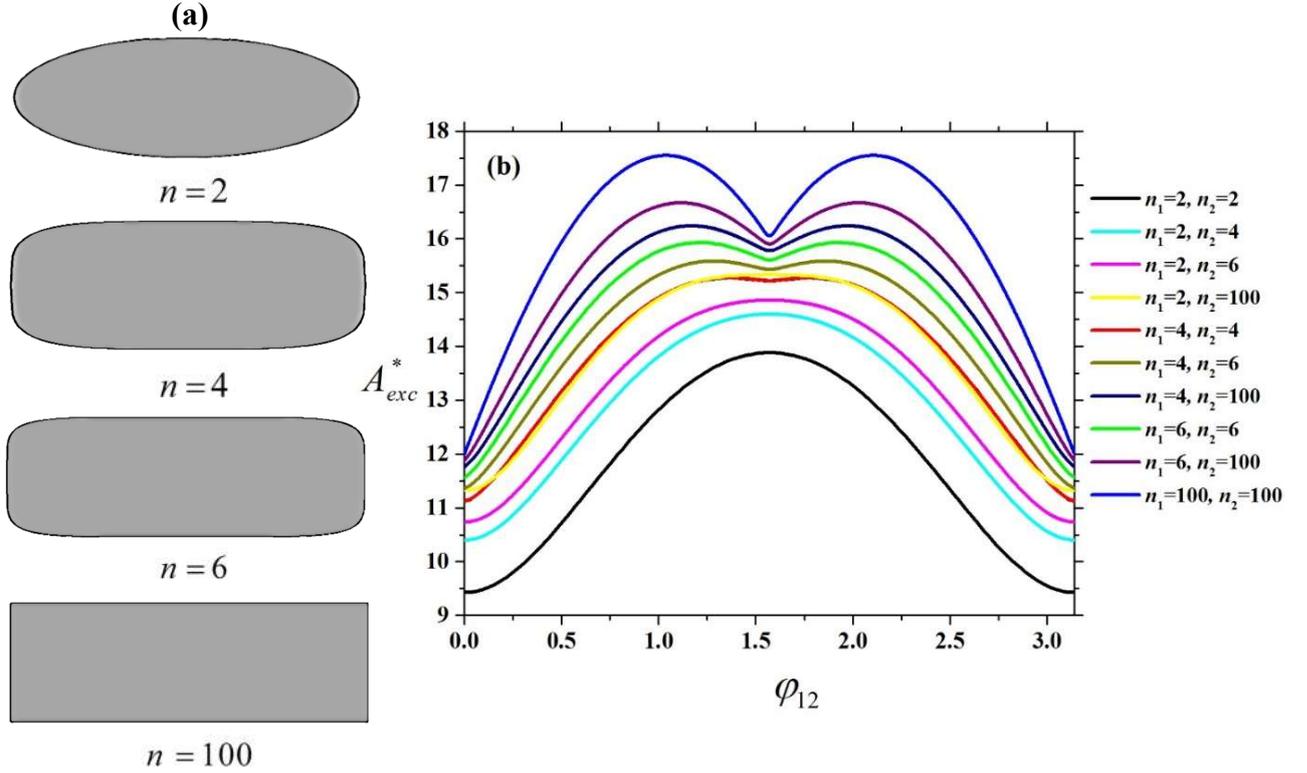

**Figure 2**: The dependence of the shape and the excluded area of hard superellipses from $n$ at an aspect ratio of $k=3$. The following cases are shown: $n=2,4,6$ and 100. (a) The effect of the varying deformation parameter ($n$) on the shape of the superellipse. (b) The excluded area between two hard superellipses with deformation parameters $n_1$ and $n_2$ as a function of angle ($\varphi_{12}$), where $\varphi_{12}$ is the angle between the major axes of two particles. All possible combinations of $n$ (2,4,6 and 100) are shown together. The length of the short ($2a$) and long ($2b$) axes of the particles are kept fixed.

An interesting system to study is the binary mixture of short and long rods which exhibit strong demixing behavior in three dimensions [50]. Fig. 3 shows the phase diagram of the binary mixture of short and long superellipses for $k_1=5$, $k_2=15$ and $d=1$ in reduced pressure-composition plane. We can see that the shape has big impact on the coexisting mole fractions at both low and high pressures. The nematic–nematic demixing transition is weakest and postponed to high pressures in the case of rectangle-rectangle mixture, while it becomes very stable if the long component is changed to be ellipse (see Fig. 3(a)). The isotropic–nematic (I-N) phase transition exhibits strange behaviour as it changes from second order to first order with increasing pressures. The first order I-N transition is very weakly fractionated for short ellipse–long rectangle mixture, while it shows very big difference in the coexisting mole fractions for short rectangle–long ellipse mixture. It can be also seen that the nematic–nematic (N–N) demixing transition terminates in a lower critical point for rectangle–rectangle, ellipse–ellipse and ellipse-rectangle mixture, while a 2$^{nd}$ order I–N transition bifurcates and a 1$^{st}$ order

N–N transition merges from a 1st order isotropic–nematic coexistence in the case of rectangle-ellipse mixture. To get further insight into the behaviour of I–N and N–N transitions, we consider the short superellipse –long ellipse mixture in Fig. 3(b), where the deformation parameter of the superellipse ($n_1$) varies between 2 and 6. It can be seen that the increasing $n_1$ widens the coexistence regions of both I-N and N-N transitions. Comparing with the results of ellipse–ellipse mixture, we can observe that a critical endpoint and an upper N–N critical point emerge for $n_1 = 4$, while the lower N–N critical point occurs at a lower pressure. At $n_1 = 6$ the lower and the upper N–N critical pressures are not present due to the stabilisation of N–N demixing transition at low pressures. The merge of upper and lower critical point takes place around $n_1 = 5$, which is not considered in this study. It is also interesting that the reentrance of the isotropic phase with increasing pressure becomes more pronounced with growing $n_1$ as the mole fraction gap of I–N transition widens with $n_1$. For example we can observe an isotropic phase which is in coexistence with a nematic one at intermediate pressures if $0.45 < x_1 < 0.82$ for $n_1 = 6$, while this phase sequence occurs only in the interval of $0.73 < x_1 < 0.83$ for $n_1 = 2$.

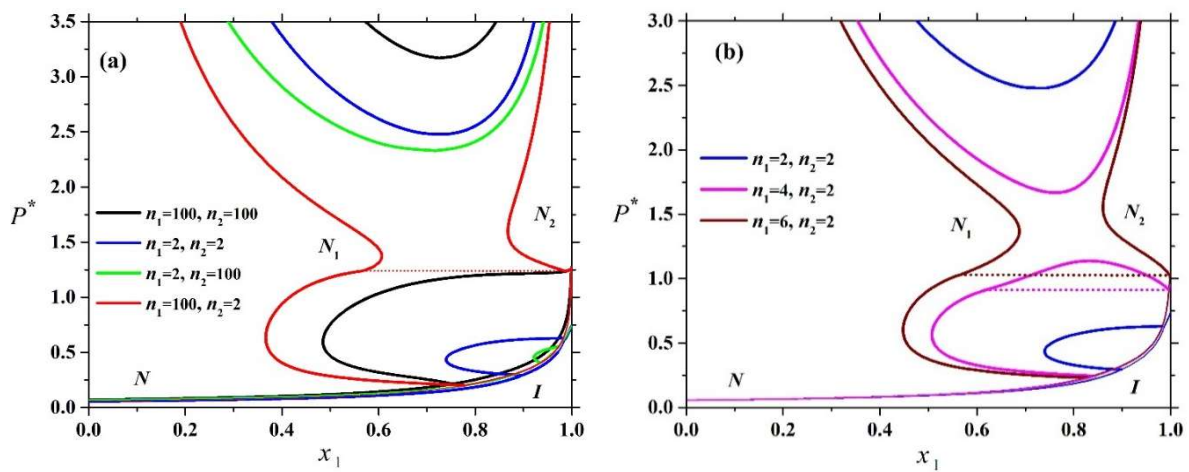

**Figure 3:** The effect of varying shape on the phase diagram of the binary mixture of short and long rods, where both components have the same short lengths ($d = 1$). The particles of the first component are short ($k_1 = 5$), while those of the second one are long ($k_2 = 15$). Figure (a) shows the phase diagrams of the rectangle–rectangle, ellipse–ellipse, ellipse–rectangle and rectangle–ellipse mixtures. In Figure (b) the long component is always ellipse, while the short one is superellipse. The phase diagram is presented in pressure-composition plane, where $x_1$ is the mole fraction of the short particles. *I* and *N* denote the isotropic and nematic phases, respectively. The dashed line represents the I-N coexistence at the critical endpoint.

In the following, our aim is to characterize the demixing tendency of the superellipse mixture with simple quantities. For this purpose it is worth mapping the binary mixture of hard superellipses into a nonadditive binary mixture of hard disks, where the nonadditivity parameter is in direct relationship with the demixing property. In more details, the binary mixture of nonadditive hard disks is described with three parameters: $\sigma_1$ and $\sigma_2$ are the diameters of the components and $\Delta$ is the nonadditivity parameter effecting only the unlike contact distance, $\sigma_{12}$, via

$$\sigma_{12} = \frac{\sigma_1 + \sigma_2}{2}(1+\Delta) . \tag{15}$$

It is shown that $\Delta$ must be high enough (>0.2) to obtain fluid-fluid demixing transition in the binary mixture of hard disks [43]. By equating the like ($A_{exc}^{11}$, $A_{exc}^{22}$) and unlike ($A_{exc}^{12}$) excluded areas of the nonadditive hard disk mixture and those of hard superellipse mixtures, we can get the corresponding molecular parameters of the binary hard disk mixture ($\sigma_1$, $\sigma_2$ and $\Delta$) in perfectly nematic (the particles are parallel) and isotropic phases. In this way we can derive that the nematic nonadditivity parameter is given by

$$\Delta_N = 2\frac{\sqrt{A_{exc}^{12}(\varphi_{12}=0)}}{\sqrt{A_{exc}^{11}(\varphi_{12}=0)} + \sqrt{A_{exc}^{22}(\varphi_{12}=0)}} - 1, \tag{16}$$

where the like and unlike excluded areas are evaluated at parallel configurations. In the isotropic phase we can get the same type of formula, but the isotropically averaged excluded areas must be used in Eq. (16), i.e.

$$\Delta_I = 2\frac{\sqrt{\langle\langle A_{exc}^{12}\rangle\rangle}}{\sqrt{\langle\langle A_{exc}^{11}\rangle\rangle} + \sqrt{\langle\langle A_{exc}^{22}\rangle\rangle}} - 1, \tag{17}$$

where $f_i(\varphi) = 1/2\pi$ has to be substituted into Eq. (4). One can show that $\Delta_N$ is still negative when the unlike excluded area equal with the geometric mean of the like excluded areas, i.e. $A_{exc}^{12} = \sqrt{A_{exc}^{11} A_{exc}^{22}}$, while it turns to be positive for the arithmetic mean ($A_{exc}^{12} = (A_{exc}^{11} + A_{exc}^{22})/2$). This shows that the positive value of $\Delta_N$ corresponds to the situation when the mixing of the components is not favorable for at least one of the components because $A_{exc}^{12} > \min(A_{exc}^{11}, A_{exc}^{22})$.

With increasing $\Delta_N$ it may happen that the mixing reduces the available room for both components, which is not favorable.

Fig. 4 illustrates the demixing tendencies of the superellipse mixtures in the isotropic and nematic phases including the systems, which are discussed in Fig. 3. It can be seen that the isotropic demixing tendency is very weak for all mixtures as $\Delta_I$ is negative or just slightly higher than zero. Opposite to this, the parallel nematic ordering has positive $\Delta_N$, which indicates that the components of the mixture do not like to be mixed together as the unlike excluded area reduces the available room for the particle in a higher extent than the like terms. In agreement with Fig. 3 we can see that the highest values of $\Delta_I$ and $\Delta_N$ belong to rectangle–ellipse mixture, which has the widest I–N and N–N coexistence regions. Fig. 4 explains also why the weak I–N transition is accompanied by a relatively strong N–N demixing transition in the ellipse–rectangle mixture (see Fig. 3). Namely, $\Delta_I$ is the smallest, while the $\Delta_N$ is almost the largest for ellipse–rectangle mixture in comparison with the other mixtures. The weakest N–N demixing is observed in the rectangle-rectangle mixture for which $\Delta_N$ is the lowest. These results show that $\Delta_N$ informs us about the N–N demixing tendency, while $\Delta_I$ is relevant for I–N transition. To find wide coexistence regions in I–N and N–N transitions we observe that both $\Delta_I$ and $\Delta_N$ should be enough high. Therefore stronger I–N and N–N transitions can be observed for systems with $d$ in the interval of $0.25 < d < 1$, because both nonadditivity parameters are higher than $\Delta$ values of the case of $d = 1$.

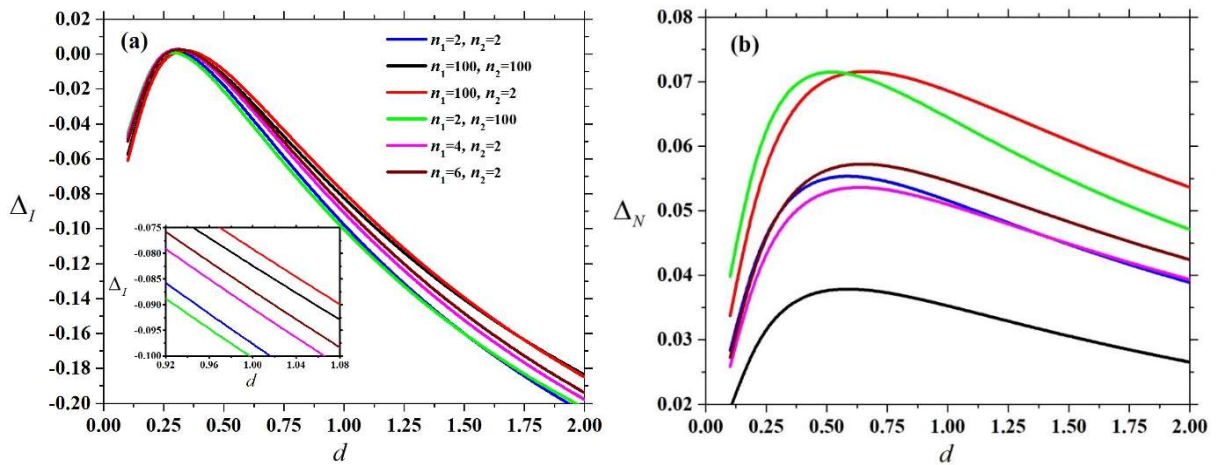

**Figure 4**: Nonadditivity parameters versus diameter ratio of some binary mixtures of hard superellipses with $k_1 = 5$ and $k_2 = 15$: (a) isotropic nonadditivity parameter, (b) nematic nonadditivity parameter. The inset enlarges a small part of (a). The color key is the same in (a) and (b).

For $d > 1$ we observe that the coexistence regions are narrower. Our search for finding the highest values of the nonadditivity parameters has resulted in the binary system of hard disks and hard needles. Fig. 5 demonstrates that lowering $k_1$ results in higher values for both $\Delta_N$ and $\Delta_I$ if the second component is almost needle. We find that the highest value of $\Delta_I$ is $\sqrt{2/(2\sqrt{2}-1)} - 1 \approx 0.046$ in the disk-needle mixture, which is not enough high to induce isotropic–isotropic (I–I) demixing transition. However, the high values of $\Delta_I$ and $\Delta_N$ give rise to very strong fractionation (wide coexistence gap) occurring between the coexisting isotropic and nematic phases. To observe I–I demixing, we believe that much higher values of $\Delta_I$ should be required, which is not possible to achieve with the present superellipse model. Therefore we are quite sure that I–I demixing transition cannot be found in binary superellipse systems using the scaled particle theory.

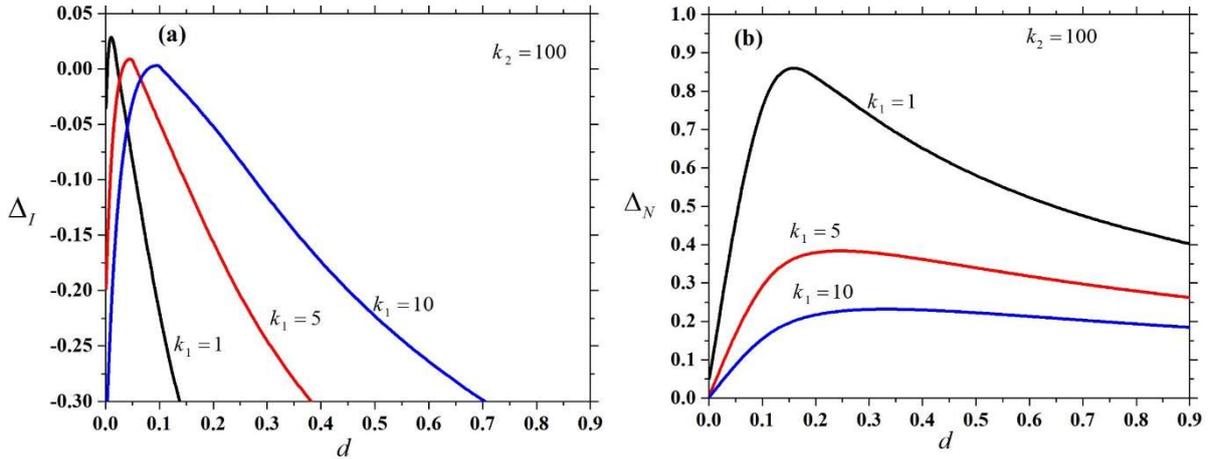

**Figure 5**: Nonadditivity parameters versus diameter ratio of some binary mixtures of superellipses where $k_2 = 100$ and $k_1 = 1, 5, 10$. (a) isotropic nonadditivity parameter, (b) nematic nonadditivity parameter. Both $n_1$ and $n_2$ are chosen to be two (ellipse–ellipse mixture).

As a demonstration we show the phase diagrams of two systems of Fig. 5. One is chosen to have high $\Delta_N$ value (Fig. 6(a)), while the other has almost maximal $\Delta_I$ value (Fig. 6(b)). We can see that $\Delta_N$ has effect on the I–N transition, too, because $\Delta_I$ is very low for the system of Fig. 6(a). Its effect manifests in very strong fractionation between the coexisting isotropic and nematic phases. The effect of $\Delta_I$ is weaker even if it is positive for the system of Fig. 6(b). The relatively high value of $\Delta_I$ is not accompanied even by the $1^{st}$ order I–N transition, because $\Delta_N$ value is low in this system. Even the N–N demixing curve is postponed to very high values of the pressure.

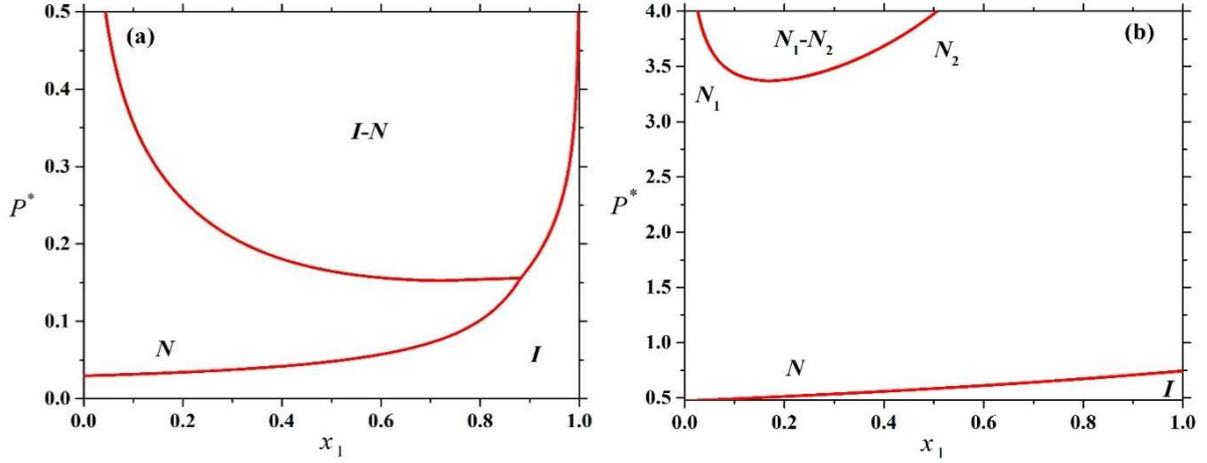

**Figure 6**: Phase diagrams of binary mixtures of hard ellipses ($n_1 = n_2 = 2$) where $k_1 = 5, k_2 = 100$. The following cases are considered: (a) $d = 0.2$. and (b) $d = 0.05$. The phase diagram is presented in pressure-composition plane. *I* and *N* denote the isotropic and nematic phases, respectively.

We have also considered the possibility of tetratic ordering in the binary mixture of hard superellipses. The tetratic phase is characterized by $S_2 = 0$ and $S_4 \neq 0$. This is different from the nematic phase, where $S_2 \neq 0$. This means that the orientational distribution function is periodic by $\pi/2$ in the case of tetratic phase, while the nematic has the period of $\pi$. It has been observed that the tetratic phase can be stable in the fluid of weakly anisotropic hard rectangles $(k \geq 1)$, including the hard squares, too [14]. However, the tetratic phase can be stable only in very dense systems, where the packing fraction is more than about 0.8 [17,18]. To find tetratic ordering at lower packing fractions we mix the hard squares with long superellipses. The orientationally ordered phase must be nematic for long particles, but it remains tetratic for the squares due to symmetry reasons. We show the phase diagram of some mixtures of squares and superellipses in Fig. 7. In all cases the effect of adding hard squares to the sea of rod-like particles destabilizes the I–N transition by shifting the I–N transition to the direction of higher pressures. This is not surprising as the squares do not form nematic phase, but they have isotropic–tetratic phase transition at $\eta \approx 0.86$ and $P^* \approx 53.6$. The destabilization effect of the squares on the I–N curves seems to be weak up to $x_1 = 0.7$ for all case, while it becomes very strong for $x_1 > 0.9$. This indicates that the squares can stay in tetratic order in a wide range of the composition. The other interesting feature of Fig. 7 is that the N–N demixing transition occurs in the square rich region of the composition ($x_1 > 0.5$). This means that the squares induces a depletion attraction between the nematically ordered superellipses. This transition can be considered as a "vapour-liquid" transition of oriented superellipses in the sea of hard squares. It is also interesting that this transition terminates in a lower critical pressure and I–N transition terminates in critical endpoint. The system which has the widest stability region of orientationally orderd phase is the square–rectangle mixture, because the N-N demixing is shifted to higher pressure values than in the other two cases. The extent of tetratic ordering of the hard squares surrounded by the nematically ordered long rectangles is presented in Fig. 8 at some pressures, where the tetratic order parameter of the squares and the corresponding packing fraction is shown as a function of mole fraction of the squares ($x_1$) up to the I–N bifurcation point. We can see that the long rods acts as an external orienting field on the squares by forcing them to be parallel or perpendicular to the nematic director. However, the strength of the orienting field increases slowly with the pressure as $S_4$ does not exceed 0.2 even at $P^* = 0.5$ (Fig. 8 (a)). We can also see that the packing fraction of the isotropic–tetratic transition of hard squares ($\eta_{IT} \approx 0.86$) is always larger than the corresponding packing fraction of the mixture at the pressures shown in Figure 8 (b). This indicates that the tetratic ordering can be stabilized at lower packing fractions in the mixture of squares and rectangles. This tetratic ordering is weak, but it can exist at low packing fractions (even at $\eta = 0.4$).

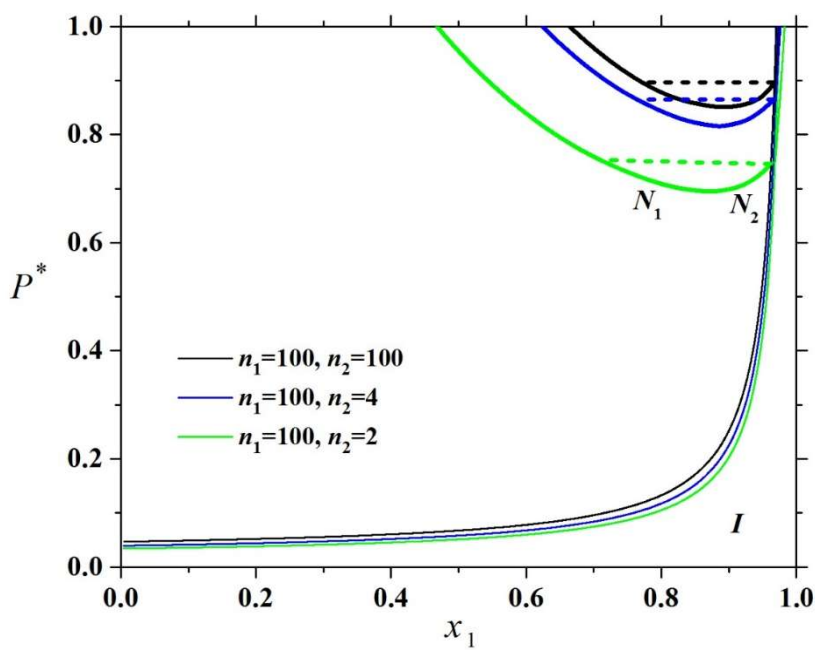

**Figure 7**: The effect of changing deformation parameter on the phase diagram of hard squares and hard superellipses. The first component is always hard square ($k_1 = 1$), while the second one is superellipse $(k_2 = 10)$. The diameter ratio (*d*) is chosen to be 2. The phase diagram is presented in pressure-composition plane. *I* and *N* denote the isotropic and nematic phases, respectively. The horizontal dashed line represents the I–N coexistence at the critical endpoint.

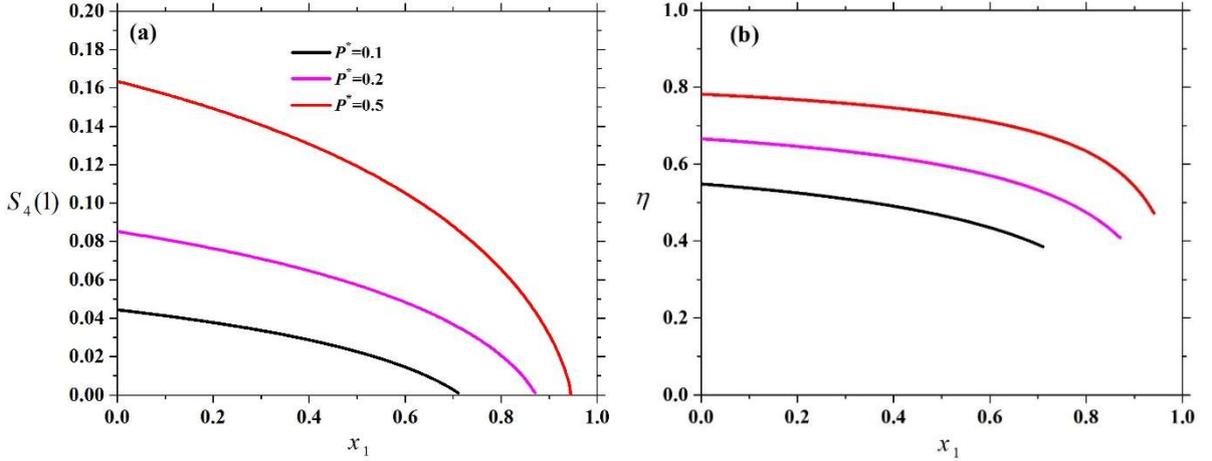

**Figure 8**: Tetratic ordering of hard squares in the mixture of hard squares and hard rectangles. Tetratic order parameter (a) and packing fraction (b) as a function of composition at some pressures, where the rectangles form nematic phase. The molecular properties of the mixture: $k_1 = 1$, $k_2 = 10$, $d = 2$ and $n_1 = n_2 = 100$.

Finally we have examined the effect of varying diameter ratio on the phase diagram, via the corresponding nonadditivity parameters. At the above examined case presented in Fig. 8 ($d = 2$) both $\Delta_I$ and $\Delta_N$ are low, which is the reason of having weak 1st order N–N demixing and 2nd order I–N transition in the mixture. As can be seen in Fig. 9 the effect of lowering $d$ is that the N-N demixing and the I–N transition become stronger, which shrinks the stability region of the nematic phase. With increasing $d$ the demixing tendency weakens, which makes a wider room for the nematic phase. However, the squares becomes even smaller with respect to the rectangles, so the tetratic ordering also weakens. Therefore the stabilization of tetratic phase at low packing fraction is feasible in such square–rectangle mixtures, where the rectangles are enough long to form nematic phase at low packing fractions and the squares are not very small with respect to the rectangles.

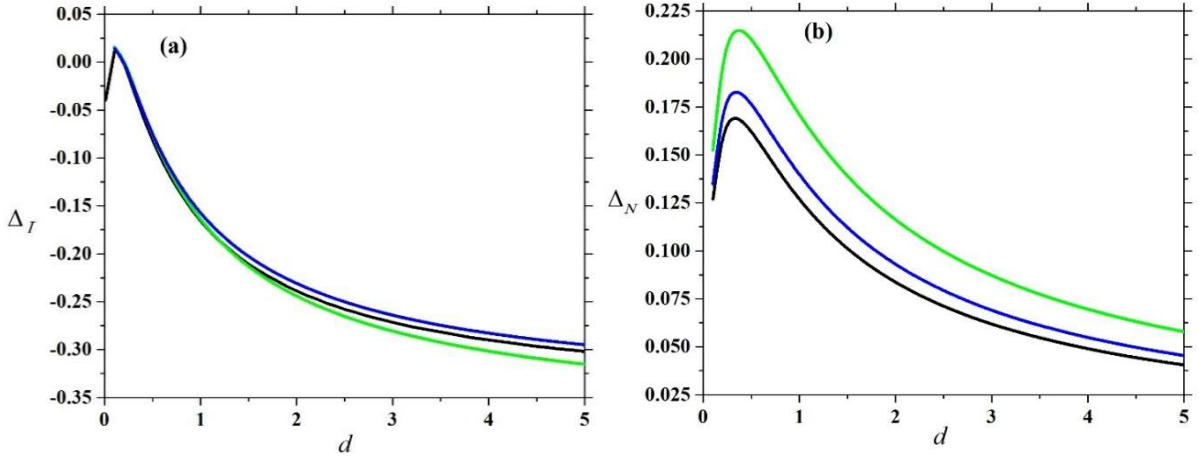

**Figure 9**: Nonadditivity parameters versus diameter ratio (*d*) of some binary mixtures of hard superellipses with $k_1 = 1$ and $k_2 = 10$ and $n_1 = n_2 = 100$: (a) isotropic nonadditivity parameter, (b) nematic nonadditivity parameter.

**Conclusion**

We have presented the results of the scaled particle theory for binary mixtures of hard superellipses, where the particles are staying in the *xy* plane. Using superellipse shape it is managed to extrapolate between the ellipse and the rectangle shapes with changing deformation parameter $n_1$ and $n_2$. This model helped us to examine the effect of size- and shape differences on the stability of isotropic and orientationally ordered (nematic and tetratic) phases. We found 1st order I–N phase transition with strong fractionation between the coexisting phases and 1st order N–N demixing transition. In addition to these transitions, a weak tetratic ordering of hard squares can be observed in the nematic host of long particles.

The observed phase diagrams are very sensitive to the shape of the constituting particles, i.e. a small change in the shape may result in a big change in the phase diagram. For example the exchanges of the short ellipses with short rectangles and the long rectangles with long ellipses result in dramatic changes in the I–N and N–N transition properties of the mixture of short and long particles even if the aspect ratios and the diameter ratio are kept fixed. The I–N and N–N coexistence regions becomes very wide in short rectangle–long ellipse mixture, while the I–N transition is only weakly fractionated and the N–N transition is postponed to the direction of higher pressures in short ellipse–long rectangle mixture. As the N–N demixing and the I-N fractionation is related to the unlike (ellipse–rectangle) interactions, our results show that the unlike excluded areas have big impact on the phase behavior of the binary mixture of short and long particles. With changing only the $n_1$ deformation parameter of the short particles we can find three types of phase diagrams: 1) weakly fractionated I–N transition and N–N demixing with lower critical point, 2) fractionated I–N transition and N–N demixing with both lower and upper critical points and 3) strongly fractionated I–N transition and N–N demixing without critical points. To explain the observed trends we mapped the binary mixture of superellipses into a binary mixture of nonadditive hard disks to obtain effective isotropic and nematic nonadditivity parameters $\Delta_I$ and $\Delta_N$. Using these parameters we can explain the observed trends in the change of the phase diagrams. To induce fractionation between the coexisting phases $\Delta_N$ must be positive, while $\Delta_I$ can be even negative. Our conclusion is that higher $\Delta_N$ and $\Delta_I$ give rise to stronger segregation between the components.

We have also looked for the possibility of a demixing transition, where the coexisting phases are isotropic. To do this we have searched for the maximum of $\Delta_I$ using a simplex algorithm. The maximum is located in the system of needle-disk mixture, where $\Delta_I = \sqrt{2/(2\sqrt{2}-1)} - 1$. This value proved to be too small to induce I–I demixing transition. However this value is enough high to produce very strong fractionation between the coexisting isotropic and nematic phases. These results are consistent with the prediction of [37].

The stability of tetratic ordering has been examined in the mixture of squares and superellipses. With adding hard squares into the sea of ordered superellipses we found that the superellipses act as a quadrupolar ordering field for the squares. The observed phase is nematic for superellipses, while it is tetratic for squares. The strength of the ordering field is very weak, so the tetratic order parameter of the squares is low. To induce stronger tetratic order, the pressure should be increased substantially. The interesting feature of the tetratic order in binary mixtures is that the minimal packing fraction can be very low in comparison with the pure systems of squares. Therefore, the mixing of long particles with squares open the window for the detection of low density tetratic ordering.

In the presented phase diagrams the deformation parameters are taken to be even numbers, although the calculations can be performed for any real numbers of $n_1 \geq 2$ and $n_2 \geq 2$. This is due to the general form of the excluded area presented in the Appendix. This allows to get a deeper insight into the effect of varying deformation parameter. Finally we note that it is also possible to determine the distance of closest approach (contact distance) between two superellipses with our method presented in the Appendix. This distance is crucial in the overlap test of Monte Carlo simulations [16,52] and in the perturbation theories of liquid crystals [53].

**ACKNOWLEDGMENTS**

H.S. and S.M. are grateful to the Research Council of Shahid Chamran University of Ahvaz for financial support (Grant No. SP98.490) and R.A. thanks Fasa University for supporting the research and providing computing facilities. S.M. also appreciates Institute of Physics and Mechatronics, University of Pannonia, Hungary. S.V. and P. G. acknowledge the financial supports of the National Research, Development, and Innovation Office (Grant No. NKFIH K124353) and Mexican CONACyT (Grant No. A1-S-9197).

**Appendix: Excluded area between two superellipses**

The equation of a standard superellipse, where the center of the body is fixed at the origin and the semi-major axis $a$ (the semi-minor axis $b$) is parallel with axis $x(y)$, can be written as

$$\Phi(x,y) = 1 \tag{A.1}$$

where

$$\Phi(x,y) = \left(\frac{|x|}{a}\right)^n + \left(\frac{|y|}{b}\right)^n. \tag{A.2}$$

Note that $n$ is the deformation parameter, which is a positive real number. A parametric representation of the above superellipse is given by

$$\vec{r}(\theta) = r(\theta)\begin{pmatrix} \cos\theta \\ \sin\theta \end{pmatrix}, \tag{A.3}$$

where

$$r(\theta) = \left[\left(\frac{|\cos\theta|}{a}\right)^n + \left(\frac{|\sin\theta|}{b}\right)^n\right]^{\frac{-1}{n}}. \tag{A.4}$$

In Eqs.(A.3) and (A.4) $r$ is the usual radial distance and $\theta$ is the polar angle. Note that the curve of the superellipse is smooth only and the gradient is well defined at all points of the superellipse for $n \geq 2$ even if the absolute values of $x$ and $y$ are used in (A.2). Therefore we deal only with $n \geq 2$ case. As the normal unit vector of the curve at any point $\vec{r}$ is given by $\vec{n}(\vec{r}) = \nabla \Phi(\vec{r}) / |\nabla \Phi(\vec{r})|$, which points out from the superellipse. Using our parametric representation, $\nabla \Phi(\vec{r})$ can be expressed as

$$\nabla \Phi(\vec{r}) = n [r(\theta)]^{n-1} \begin{pmatrix} \dfrac{|\cos\theta|^{n-2}}{a^n} \cos\theta \\ \dfrac{|\sin\theta|^{n-2}}{b^n} \sin\theta \end{pmatrix}. \quad (A.5)$$

Let us continue with two superellipses, which are characterized by $a_1, b_1, n_1$ and $a_2, b_2, n_2$ parameters and $\varphi_{12}$ is the angle between their main axes. We use the notations of Fig. (A), where $\vec{r}_1$ and $\vec{r}_2$ position vectors points to the common point of the two bodies, when they are in contact. The polar angles of these vectors are denoted by $\theta_1$ and $\theta_2$, which are measured from $x$ and $x'$ axes of the respective body fixed coordinate systems. The angle between the two coordinate systems ($\varphi_{12}$) is the angle between the two superellipses. In the following derivations, we determine the excluded areas for smooth superellipes only, i.e. $n_1, n_2 \geq 2$. Fig. (A) shows that the two curves have common tangent at the contact point. The normal vectors of both superellipses point outside from the corresponding curves and must have exactly the opposite directions, i.e. $\vec{n}_1(\vec{r}_1) = -\vec{n}_2(\vec{r}_2)$, because the particles touch each other from outside. As the particle 2 is rotated by $\varphi_{12}$ angle with respect to the particle 1, its normal vector is also rotated by $\varphi_{12}$. Therefore the condition for the normal vectors becomes

$$\frac{\vec{\nabla}\Phi_1(\vec{r}_1(\theta_1))}{|\vec{\nabla}\Phi_1(\vec{r}_1(\theta_1))|} = -\frac{\hat{R}(\varphi_{12}) \vec{\nabla}\Phi_2(\vec{r}_2(\theta_2))}{|\vec{\nabla}\Phi_2(\vec{r}_2(\theta_2))|}, \quad (A.6)$$

where the rotational matrix $\hat{R}(\varphi_{12})$ is given by

$$\hat{R}(\varphi_{12}) = \begin{pmatrix} \cos\varphi_{12} & -\sin\varphi_{12} \\ \sin\varphi_{12} & \cos\varphi_{12} \end{pmatrix}. \quad (A.7)$$

Note that the gradients are taken in the particle's fixed frames and that the rotation does not affect the norm of the vector. Substituting Eqs. (A.4) and (A.5) with the corresponding indices of the particles into Eq. (6) and using the rotational matrix Eq. (A.7), we get the following two equations for the $x$ and $y$ components of Eq. (A.6)

$$\frac{\cos\theta_1 |\cos\theta_1|^{n_1-2}}{\left[|\cos\theta_1|^{2(n_1-1)} + \left(\frac{a_1}{b_1}\right)^{2n_1} |\sin\theta_1|^{2(n_1-1)}\right]^{\frac{1}{2}}} = \frac{-\cos\varphi_{12}\cos\theta_2 |\cos\theta_2|^{n_2-2} + \left(\frac{a_2}{b_2}\right)^{n_2}\sin\varphi_{12}\sin\theta_2 |\sin\theta_2|^{n_2-2}}{\left[|\cos\theta_2|^{2(n_2-1)} + \left(\frac{a_2}{b_2}\right)^{2n_2} |\sin\theta_2|^{2(n_2-1)}\right]^{\frac{1}{2}}},$$

(A.8)

and

$$\frac{\sin\theta_1 |\sin\theta_1|^{n_1-2}}{\left[\left(\frac{b_1}{a_1}\right)^{2n_1}|\cos\theta_1|^{2(n_1-1)} + |\sin\theta_1|^{2(n_1-1)}\right]^{\frac{1}{2}}} = \frac{-\sin\varphi_{12}\cos\theta_2 |\cos\theta_2|^{n_2-2} - \left(\frac{a_2}{b_2}\right)^{n_2}\cos\varphi_{12}\sin\theta_2 |\sin\theta_2|^{n_2-2}}{\left[|\cos\theta_2|^{2(n_2-1)} + \left(\frac{a_2}{b_2}\right)^{2n_2} |\sin\theta_2|^{2(n_2-1)}\right]^{\frac{1}{2}}}.$$

(A.9)

Eqs. (A.8) and (A.9) provide a relationship between $\theta_1$ and $\theta_2$ when the two particles are in contact. This means practically that, for a given relative orientation $\varphi_{12}$, supposing that the second particle touch the first one at $\vec{r}_2(\theta_2)$ point, the simultaneous solution of Eqs. (A.8) and (A.9) provides $\theta_1$ as a function $\theta_2$ and $\vec{r}_1(\theta_1)$ points to the contact point of the particle 1. The contact distance vector $\vec{\sigma}$, which connects the centers of the two particles when they are in contact, depends only on $\theta_2$, because $\vec{\sigma} = \vec{r}_1 - \vec{r}_2$. Therefore, the excluded area between the first and second particles can be expressed from the well-known formula for the area of two-dimensional objects as

$$A_{exc} = \frac{1}{2}\int_0^{2\pi} \hat{e}_z \left(\vec{\sigma} \times \frac{d\vec{\sigma}}{d\theta_2}\right) d\theta_2 \qquad (A.10)$$

where $\hat{e}_z$ is the unit vector perpendicular to the $xy$ plane, which selects the $z$ component of the cross product. Using $\vec{\sigma} = \vec{r}_1 - \vec{r}_2$ and the identity of $-\vec{r}_1 \times \frac{d\vec{r}_2}{d\theta_2} - \vec{r}_2 \times \frac{d\vec{r}_1}{d\theta_2} = \frac{d(\vec{r}_1 \times \vec{r}_2)}{d\theta_2} + 2\frac{d\vec{r}_2}{d\theta_2} \times \vec{r}_1$ we can rewrite Eq (A. 10) as

$$A_{exc} = \tilde{a}_1 + \tilde{a}_2 + \int_0^{2\pi} \hat{e}_z \left( \frac{d\vec{r}_2}{d\theta_2} \times \vec{r}_1 \right) d\theta_2 \qquad (A11)$$

where

$$\tilde{a}_i = \frac{1}{2} \int_0^{2\pi} \hat{e}_z \left( \vec{r}_i \times \frac{d\vec{r}_i}{d\theta_2} \right) d\theta_2$$

$$= \frac{a_i b_i}{n_i} 2^{2\left(1-\frac{1}{n_i}\right)} \sqrt{\pi} \frac{\Gamma\left(\frac{1}{n_i}\right)}{\Gamma\left(\frac{1}{n_i} + \frac{1}{2}\right)} \qquad (A.12)$$

is the area of particle *i*. Note that in the derivation of Eq. (A.11) we have performed an integration using the integration by parts method. To compute the last term of Eq. (A.11) we need the components of $\vec{r}_1$ and $\vec{r}_2$ relative to the same coordinate frame. Using the coordinate frame, which is fixed to the particle 1, the derivate in Eq. (A11) is given by

$$\frac{d\vec{r}_2}{d\theta_2} = \hat{R}(\varphi_{12}) \left[ \frac{dr_2}{d\theta_2} \begin{pmatrix} \cos\theta_2 \\ \sin\theta_2 \end{pmatrix} + r_2 \begin{pmatrix} -\sin\theta_2 \\ \cos\theta_2 \end{pmatrix} \right] . \qquad (A.13)$$

Using this equation, the final form of the excluded area can be written as

$$A_{exc} = \tilde{a}_1 + \tilde{a}_2 + \int_0^{2\pi} d\theta_2 \left( \left[ \frac{dr_2}{d\theta_2} \cos(\theta_2 + \varphi_{12}) - r_2 \sin(\theta_2 + \varphi_{12}) \right] \sin(\theta_1) \right.$$
$$\left. - \left[ \frac{dr_2}{d\theta_2} \sin(\theta_2 + \varphi_{12}) + r_2 \cos(\theta_2 + \varphi_{12}) \right] \cos(\theta_1) \right) r_1, \qquad (A.14)$$

Here we used the central symmetry of the excluded area, therefore the integral over $\theta_2$ runs only up to $\pi$. In the above expression $\theta_1$, $r_1$ and $r_2$ are considered to be functions of $\theta_2$ from the solution of Eqs. (A.8)-(A.9). Note that we need only $\sin\theta_1$ and $\cos\theta_1$ as functions of $\theta_2$ in the excluded area calculations (A.14). Now we show that it is possible to get analytic formulas for $\sin\theta_1$ and $\cos\theta_1$ in the following way. Let us denote the r.h.s of Eqs. (A.8) and (A.9) by *F* and *G*, respectively, which functions depend on the molecular parameters of particle 2, $\theta_2$ and $\varphi_{12}$. Using *F* we get from Eq. (A. 8) that

$$\left(\cos^2\theta_1\right)^{n_1-1} = \left[\left(\cos^2\theta_1\right)^{n_1-1} + \left(\frac{a_1}{b_1}\right)^{2n_1}\left(1-\cos^2\theta_1\right)^{n_1-1}\right]F^2, \tag{A.15}$$

which can be written shortly as

$$\cos^2\theta_1 = \frac{c}{1+c}, \tag{A.16}$$

where

$$c = \left[\frac{F^2}{1-F^2}\left(\frac{a_1}{b_1}\right)^{2n_1}\right]^{\frac{1}{n_1-1}}. \tag{A.17}$$

From Eq. (A.8) it is clear that the sign of $\cos\theta_1$ is the same as the sign of $F$, therefore

$$\cos\theta_1 = \text{sgn}(F)\sqrt{\frac{c}{1+c}}. \tag{A.18}$$

In a similar way from Eq. (A.9) we get that

$$\sin\theta_1 = \text{sgn}(G)\sqrt{\frac{b}{1+b}}, \tag{A.19}$$

where

$$b = \left[\frac{G^2}{1-G^2}\left(\frac{b_1}{a_1}\right)^{2n_1}\right]^{\frac{1}{n_1-1}}. \tag{A.20}$$

We must mention here that Eqs. (A.18) and (A.19) are consistent with each other satisfying the identity $\cos^2\theta_1 + \sin^2\theta_1 = 1$. Substituting Eqs. (A.18) and (A.19) into Eq. (A.14) the excluded area is expressed explicitly as a function of $\theta_2$. Using these tricks the integrals can be performed easily using a simple numerical quadrature without resorting to equation solvers for Eqs. (A.8) and (A.9).

By the appropriate choice of the molecular parameters Eq. (A14) can provide the like $\left(A_{exc}^{11}, A_{exc}^{22}\right)$ and the unlike ($A_{exc}^{12}, A_{exc}^{21}$) excluded areas. If particle 1 and the particle 2 are identical, i.e. both particles are characterized with $(a_1, b_1, n_1)$ or $(a_2, b_2, n_2)$ we get $A_{exc}^{11}$ or $A_{exc}^{22}$

However we get $A_{exc}^{21}$ and $A_{exc}^{12}$ if the particles have different parameters, i.e. $(a_1, b_1, n_1) \neq (a_2, b_2, n_2)$.

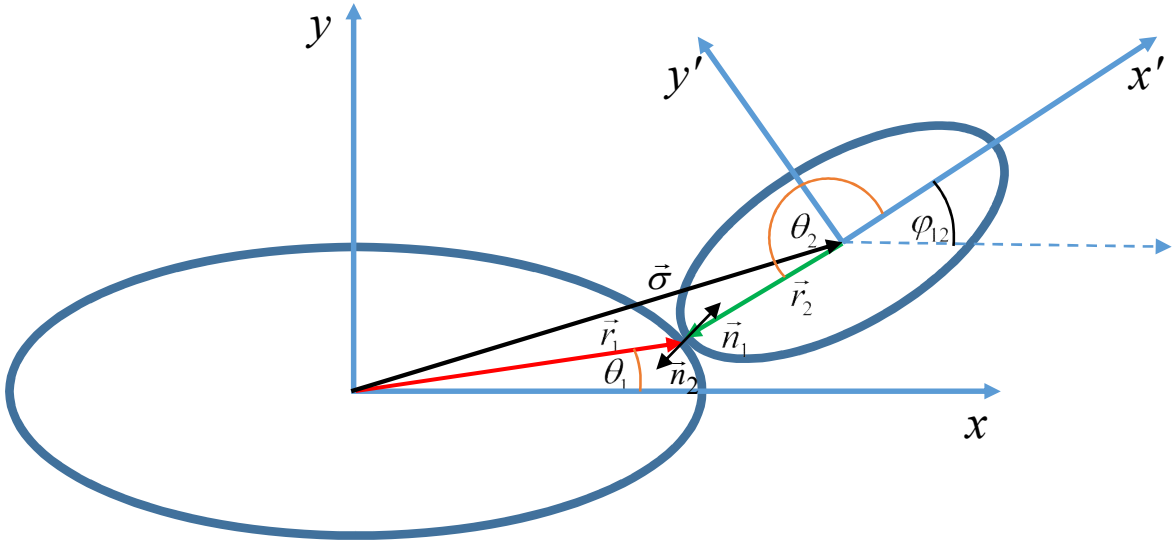

**Figure A**: Two superellipses are in contact externally. The following notation is used: the angle between the major axes of the two superellipses is denoted by $\varphi_{12}$, $\vec{\sigma}$ is the contact distance vector between the two superellipses, $\theta_1$ and $\theta_2$ are the orientations of the position vectors $\vec{r}_1$ and $\vec{r}_2$ of the contact point which are measured in the coordinate frame fixed to particle 1 and 2, respectively. Moreover $\vec{n}_1$ and $\vec{n}_2$ are the normal vectors of the common tangent.

## Data Availability Statement

The data that support the findings of this study are available from the corresponding author upon reasonable request.